\documentclass{article}

\usepackage[english]{babel}
\usepackage{amssymb,physics}
\usepackage{float}
\usepackage[caption = false]{subfig}
\usepackage[letterpaper,top=2cm,bottom=2cm,left=3cm,right=3cm,marginparwidth=1.75cm]{geometry}
\usepackage{authblk}
\usepackage{amsmath}
\usepackage{graphicx}
\usepackage[colorlinks=true, allcolors=blue]{hyperref}

\title{Shape Invariant Potentials in Supersymmetric Quantum Cosmology}

\author[1]{S. Jalalzadeh\thanks{shahram.jalalzadeh@ufpe.br}}

\author[2,3]{ S. M. M. Rasouli\thanks{mrasouli@ubi.pt} }
\author[2]{Paulo Moniz\thanks{pmoniz@ubi.pt}}

\affil[1]{Departamento de F\'{i}sica, Universidade Federal de Pernambuco, Recife, PE, 52171-900, Brazil}
\affil[2]{Departamento de F\'{i}sica,
Centro de Matem\'{a}tica e Aplica\c{c}\~{o}es (CMA-UBI),
Universidade da Beira Interior,
Rua Marqu\^{e}s d'Avila
e Bolama, 6200-001 Covilh\~{a}, Portugal}
\affil[3]{Department of Physics, Qazvin Branch, Islamic Azad University, Qazvin, Iran}

\begin{document}
\maketitle

\begin{abstract}
In this brief review, we comment on the concept of shape invariant potentials, which is an essential feature in many settings of $N=2$ supersymmetric quantum mechanics. To motivate its application within supersymmetric quantum cosmology, we present a case study to illustrate the value of this promising tool. Concretely, we take a spatially flat FRW model in the presence of a single scalar field, 
minimally coupled to gravity. Then, we extract the associated Schr\"odinger--Wheeler--DeWitt equation, allowing for a particular scope
of factor ordering. Subsequently, we
compute the corresponding supersymmetric partner Hamiltonians,
$H_1$ and $H_2$. Moreover, we point out how the shape invariance property can be employed to bring a relation among several factor orderings choices for our Schr\"odinger--Wheeler--DeWitt equation. The ground state is retrieved, and the excited states easily written. Finally, the Hamiltonians, $H_1$ and $H_2$ are explicitly presented within a $N=2$ 
supersymmetric quantum mechanics framework.
\end{abstract}

\section{Introduction}
\label{int}
\indent

%A short intro............

{Shape Invariant Potentials (SIP) constitute one of the hallmarks of supersymmetric quantum mechanics (SQM), in the 
sense that it enables a prolific framework to be elaborated. 
Being more specific, the presence of SIP allows to easily obtain 
the set of states for a class of quantum systems, suitably based on an elegant algebraic construction.} 
%Let us therefore begin mentioning that there exist an algebraic structure associated with the SIP framework, which has gradually been acquiring  a twofold relevance.    Concretely, it acquired a  paramount importance within   most of exactly solvable  problems in quantum mechanics 
Hence, let us begin by mentioning that there is an algebraic structure associated with the SIP framework. {It}
has gradually been acquiring a twofold relevance
%. Concretely, it acquired paramount importance 
{and} within   most of the exactly solvable  problems in quantum mechanics 
 \cite{Cooper-1,yaya,Bala,yaya-1,Chen,Khare,Quesne}. 
 
 On the one hand, such a structure has
 provided a method to determine  eigenvalues and eigenfunctions, by means of which
 a spectrum is generated.
 %Let us be 
 {More specifically, a}  broad 
 {set} of those exactly solvable cases can be assembled and assigned 
 within concrete {classes}; very few
exceptions are known \cite{Cooper-1,yaya,Bala,yaya-1,Chen,Khare,Quesne}. 
The distinguishing  feature of any of such classes is that any exactly
 solvable case bears a \textit{shape invariant potential:  supersymmetric partners} are of the same shape, and their spectra can be determined entirely by an algebraic procedure comparable to that of the harmonic oscillator.
{In other words,}   operators can be defined,
namely $A:=\frac{d}{dx}+W(x)$, 
and its Hermitian conjugate $A^\dagger :=-\frac{d}{dx}+W(x)$, 
Hamiltonians $H_1$ and its superpartner $H_2$ being 
expressed as  $A^\dagger A$ and $AA^\dagger$,  respectively. 
From this, we can produce and  operate with other (more 
adequate) ladder operators for  
correspondingly appropriate quantum
numbers. {These}  can be maneuvered within a
$J_\pm, J_3$ algebra,  with comparable 
features  to
textbook ladder operators of angular momentum within 
either
$SU(2)$ or $SO(3)$; please see \cite{Cooper-1,yaya,Bala,yaya-1,Chen,Khare,Quesne} for relevant details.

On the other hand, several of these exactly solvable systems  also possess a \textit{potential algebra}:  the corresponding Hamiltonian 
can be written as a Casimir operator of an underlying 
algebra, 
{which in particular cases is  of} a $SO(2,1)$ 
nature \cite{yaya}. 
Remarkably, there is a close correspondence:  shape invariance can be 
expressed as constraint, which assists in establishing the spectrum; moreover, this  shape invariance constraint 
can be written  as an algebraic condition. For a specific set of SIP, the algebraic  condition
corresponds to  the mentioned $SO(2,1)$ potential algebra, where 
unitary representations become of crucial use. {Interestingly}, 
these can be related to that of  $SO(3)$; several SIP’s are as such \cite{Cooper-1,yaya,Bala,yaya-1,Chen,Khare,Quesne}.  On the 
whole, a connection between SIP  and \textit{potential algebra} was attained. {Nevertheless}, it is also clear that 
in spite of the structural similarity between $SO(2,1)$
and $SO(3)$ algebras, there are caveats to be {aware,  related to the}  differences between those unitary representations.

There are also a couple of additional  points that we would like to 
emphasize. To start with,  
{some} quantum states
 can be retrieved  by group theoretical  methods. {This is}
 further endorsed 
{from the} 
%by those   established   
connections between shape
invariance and potential algebra, 
wherein  the former is translated into a concrete  
formulation within the latter \cite{yaya}. 
{As a result,}  the scope of the class of potentials 
where that could be applied was  made more prominent 
 \cite{Cooper-1}. Furthermore,  other classes 
 {have been explored,}
related to harmonic  oscillator {induced} second  order {differential equations,
bearing   group} and algebra features,
{which subsequently allowed more SIP   to be  found}  
\cite{Oikonomou,Cooper-1,yaya,Bala,yaya-1,Chen,Khare,Quesne,Bazeia,Stahlhofen,Fakhri,polo}. 
In particular,  this 
%concept 
was {further} extended toward graded algebras in \cite{Oikonomou}.
 
Secondly, these algebraic/group theory procedures {(within
 the concrete use of algebras such as $SO(2,1)$ or $SO(3)$)}, 
 {have} a striking 
 resemblance to the approach and descriptive language used in  \cite{Shahram-1,Shahram-2,Shahram-3,Shahram-4}; a review of this idea is found in \cite{Shahram-5}. Therein, it was pointed that   intertwining boundary conditions, the algebra of constraints and hidden symmetries in quantum cosmology 
 could be quite fruitful. Specifically,  group/algebraic  properties within  ladder operators, either from angular momentum  or from 
 within the explicit presence of specific matter fields 
 (and their properties), 
determined a partition of wave functions and boundary conditions, {according} to the Bargmann index \cite{Shahram-1,Shahram-2,Shahram-3,Shahram-4}. Moreover, 
 we proposed in \cite{Shahram-5} to extend this framework towards  SIP, which could include well known analytically solvable cosmological cases. 
{ Being more clear, provided we identify integrability in terms of the shape invariance conditions, we could eventually import those specific features of SQM 
towards quantum cosmology 
 \cite{Moniz-1,Moniz-2}. } That was the challenge we laid out in \cite{Shahram-5}, which is still to be addressed:  we hope our review paper herein can further enthuse 
 {someone to pick this up}.  
 A somewhat related and interesting {direction to explore is} to also consider 
 %for that aim 
 {an  elaboration following 
 \cite{Cordero-1,Cordero-2}. Specifically,  
 %adjoined within  any suitable variant to 
 accommodating the 
 lines} in \cite{Shahram-1,Shahram-2,Shahram-3,Shahram-4,Shahram-5} plus 
 supersymmetry (SUSY) \cite{Moniz-1,Moniz-2}. In brief, this  paragraph conveys  what is our {central} motivation to produce this review, building from a suggestion advanced in \cite{Shahram-5}.

%\cite{DKS88,K97,CKS95,G83,CGK87,S40,S41,IH51,D1882,LP86,N84,P86,K77,GL51,AM80,DT93,DOT93,B93,JRA20,FA93,FR01,Oikonomou,Cooper-1,yaya,Bala,yaya-1,Chen,Khare,Quesne,Bazeia,Stahlhofen,Fakhri,polo}

And thirdly, the interest {in the above} elements notwithstanding, there are still obstacles that ought to be frankly mentioned. Namely, about the scope of the usefulness of SIP in quantum cosmology. In fact, the list of SIP is quite {restrictive,}
%intrinsic, 
and most potentials therein do not 
emerge naturally within a minisuperspace. A few do, but for very particular case studies \cite{diaz2019supersymmetric}. 
%We point to a case and elaborate herewith in our review. 
{The} classification of spatial geometries upon the Bianchi method implies that 
the potentials extracted from the 
gravitational degrees of freedom are very specific.
{Any} 'broadness' can be introduced by inserting ($i$) very specific matter fields into the minisuperspace (and therein we ought to be using realistic potentials {(as indicated by particle physics))} or instead, ($ii$) try more SIP fitted choices but at the price of being very much \textit{ad hoc} i.e., an artificial selection. Nevertheless,
the list of SIP and similar cases where the algebraic tools could be adopted, has been 
%explored and subsequently 
{extended}. Although not a strong positive endorsement, there is work \cite{2005, 2011, 2001, GANGOPADHYAYA2020126722, sym4030452, Cari_ena_2000, 10.4303/jpm/P110502,doi:10.1007/978-1-4020-5796-0,nasuda2021non,gangopadhyaya2017supersymmetric} that allow us to consider that eventually an 
extended notion of SIP may be soundly established, such that 
more cosmological minisuperspaces can be discussed within (see e.g., \cite{diaz2019supersymmetric}). For the moment, this is a purpose set in construction and where this review paper aims to to enthuse and promote towards.

%\cite{2005, 2011, 2001, GANGOPADHYAYA2020126722, sym4030452, Cari_ena_2000, 10.4303/jpm/P110502,doi:10.1007/978-1-4020-5796-0,nasuda2021non,gangopadhyaya2017supersymmetric,acosta2012algebraic}

%a few  technicalities about it will be presented in subsection \ref{SIP}. Upon that, we proceed into 
%\bl{applying it to}
%discussing 
%a specific quantum cosmological setting.
%by means of such a tool, 
%\bl{We hope that the discussion herewith
%We, therefore, hope to enthuse any reader from herewith.
%and we therefore hope to enthuse from herewith.}

%In this paper, we import a 
%very 
%\bl{particular} feature which is 
%frequently used within supersymmetric quantum mechanics (SQM), %namely Shape Invariant Potentials (SIP); will enthuse 
%researchers towards}
%which 

Upon this introductory section, this review is structured as follows. In section 2, we summarize 
%some of the 
{features of SQM that we will be 
%mentioning and sometimes 
employing.  In particular, a few  technicalities about SIP will be presented in subsection \ref{SIP}}. Then, in section 3, we take a 
 case study, typically a toy model, by means of which we aim to promote work in SUSY quantum cosmology 
 {with 
 the novel perspective of SIP. We emphasize that this 
is 
a line of investigation that has not yet been attempted before 
(see subsection \ref{SUSY-QC} for details).} 
%concept and context. 
Section 4 conveys a Discussion and suggestions of Outlook for future research work.

\section{Supersymmetric Quantum Mechanics}
\label{SUSY-QM}
\indent

In this section, let us present a brief review of some of the 
pillars that characterize SQM. Then, we proceed to add a summary of the shape invariance concept.
This section contains neither new results nor any innovated procedure, but only a very {short} overview of the results presented within seminal papers, e.g.,  \cite{CGK87,DKS88,CKS95}.

\subsection{Hamiltonian formulation of supersymmetric quantum mechanics}
\label{Ham-SUSY-QM}
\indent

In order to describe SQM, let us start with the Schr\"{o}dinger equation:
\begin{eqnarray}
 H\Psi_n(x)=\left[-\frac{\hbar^2}{2m}\frac{d^2}{dx^2}+V(x)\right]
 \Psi_n(x)=E_n \Psi_n(x).
\label{SE-GS1111}
\end{eqnarray}
We assume that the ground state wave function $\Psi_0(x)$
(that has no nodes\footnote{In \cite{CKS95}, the excited wave functions have also been studied.
More concretely, instead of the choice of a nonsingular superpotential that is based on the
ground state wave function $\Psi_0(x)$, a generalized procedure was
presented to construct all possible superpotentials.})
associated with a potential $V_1(x)$ is known. Then, by assuming that the
ground state energy $E_0$ to be zero, the Schr\"{o}dinger equation for this ground state reduces to
\begin{eqnarray}
 H_1\Psi_0(x)=\left[-\frac{\hbar^2}{2m}\frac{d^2}{dx^2}+V_1(x)\right]\Psi_0(x)=0,
\label{SE-GS}
\end{eqnarray}
which leads to construct the potential $ V_1(x)$:
\begin{eqnarray}
 V_1(x)=\frac{\hbar^2}{2m}\frac{\Psi''_0(x)}{\Psi_0(x)},
\label{V1}
\end{eqnarray}
where a prime denotes differentiation with respect to $x$.
{It} is straightforward to factorize the Hamiltonian from the operators as follows, 
\begin{eqnarray}
 A=\frac{\hbar}{\sqrt{2m}}\left[\frac{d}{dx}-\frac{\Psi'_0(x)}{\Psi_0(x)}\right], \hspace{10mm}
A^\dag=-\frac{\hbar}{\sqrt{2m}}\left[\frac{d}{dx}+\frac{\Psi'_0(x)}{\Psi_0(x)}\right],
\label{A-vs-Psi}
\end{eqnarray}
{through the } equation
\begin{eqnarray}
H_1= A^\dag A.
\label{H1}
\end{eqnarray}
In order to construct the SUSY theory related to the original Hamiltonian $H_1$, the next step is to
define another operator by reversing the order of $A$
and $A^\dag$, i.e., $H_2\equiv AA^\dag$, by which we indeed
get {the} Hamiltonian corresponding to a new potential $V_2(x)$:
\begin{eqnarray}
 H_2=-\frac{\hbar^2}{2m}\frac{d^2}{dx^2}+V_2(x),
\label{H2}
\end{eqnarray}
where
\begin{eqnarray}
 V_2(x)=-V_1(x)+\frac{\hbar}{{m}}\left[\frac{\Psi'_0(x)}{\Psi_0(x)}\right]^2.
\label{V2}
\end{eqnarray}
The potentials $V_1(x)$ and $V_2(x)$ have been
referred to as the supersymmetric partner potentials.
It should be noted that $H_2$, as the partner
Hamiltonian corresponding to $H_1$, is
in general not unique, but there
is a class of Hamiltonians {$H^{(m)}$ that can be partner Hamiltonians}\footnote{{Cf. next subsection, concretely about Eq.(\ref{sha7})}.}.
This point has been specified \cite{N84,CGK87,P86},
{conveying to a better understanding} of the 
relationship between SQM and the
inverse scattering method,  established by Gelfand and Levitan \cite{GL51,AM80}.

In SQM, instead of the
ground state wave function $\Psi_0(x)$ associated
with $H_1$, the superpotential $W(x)$ {is introduced, 
being related } to $\Psi_0(x)$ and its first derivative (with respect to $x$) {by means of}
\begin{eqnarray}
 W(x)=-\frac{\hbar}{\sqrt{2m}}\left(\frac{\Psi'_0(x)}{\Psi_0(x)}\right).
\label{W}
\end{eqnarray}
{At this stage}, it is worthy to express the operators $A$ and $A^\dag$ and the
supersymmetric partner potentials $V_1(x)$ and $ V_2(x)$ in terms of the superpotential.
Therefore, using \eqref{W}, equations \eqref{V1}, \eqref{A-vs-Psi} and \eqref{V2} are rewritten as
\begin{eqnarray}\label{A-vs-W}
 A&=&\frac{\hbar}{\sqrt{2m}}\frac{d}{dx}+W(x), \hspace{18mm}
A^\dag=-\frac{\hbar}{\sqrt{2m}}\frac{d}{dx}+W(x),\\
V_1(x)&=&-\frac{\hbar}{\sqrt{2m}}W'(x)+W^2(x), \hspace{10mm}
V_2(x)=\frac{\hbar}{\sqrt{2m}}W'(x)+W^2(x),
\label{V-vs-W}
\end{eqnarray}
where {the expressions for $V_1$ and $V_2$ in 
\eqref{V-vs-W} constitute   Riccati equations}. Equations \eqref{A-vs-W} and \eqref{V-vs-W}
imply {that} 
$W'(x)$ is proportional to the commutator of the operators
$A$ and $A^\dag$, and $W^2(x)$ is the average of the partner potentials. 

\begin{figure}
\centering\includegraphics[width=3.6in]{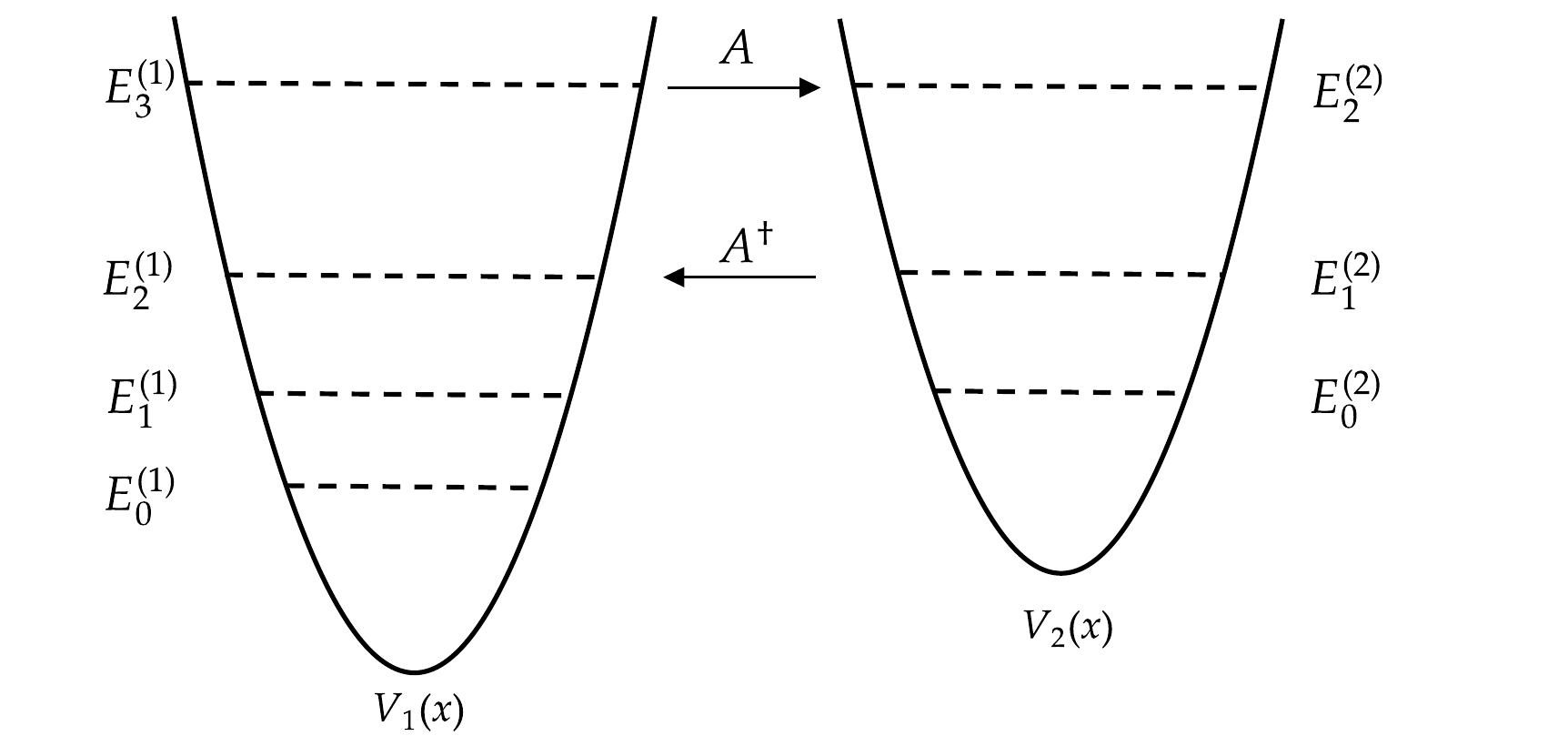}
\caption{The energy levels of $V_1(x)$ and $ V_2(x)$ as
two supersymmetric partner potentials. The figure is associated with unbroken SUSY.
It is seen that, except an extra state $E_0^{(1)}=0$, the other energy levels are degenerate.
Moreover, in this figure, it is shown that how the operators $A$ and $A^\dag$ connect eigenfunctions.
}
\label{fig1}
\end{figure}

It is easy to show that the wave functions, the energy eigenvalues
and the S-matrices of both the Hamiltonians $H_1$ and $H_2$ are related.
%In this work, let 
{Let} us merely outline the results and abstain from proving them.
In this regard, we take $\Psi_n^{(1)}$ and $\Psi_n^{(2)}$ as the
eigenfunctions of $H_1$ and {$H_2$}, respectively. 
Moreover, we denote their corresponding energy eigenvalues
with $E_n^{(1)}\geq0$ and $E_n^{(2)}\geq0$ where $n=0,1,2,3,...,$ is the
number of the nodes in the wave function. It is straightforward to
show that supersymmetric partner potentials $V_1(x)$ and $ V_2(x)$ possess
the same energy spectrum. However, we should note
that for the ground state energy $E_0^{(1)}=0$ associated
with the potential $V_1(x)$, there is no corresponding level for its partner $ V_2(x)$.
Concretely, it has been shown that
\begin{eqnarray}\label{same-1}
 E_n^{(2)}&=&E_{n+1}^{(1)}, \hspace{25mm}E_0^{(1)}=0,\\
 \label{same-2}
 \Psi_n^{(2)}&=&\left[E_{n+1}^{(1)}\right]^{-\frac{1}{2}}A\Psi_{n+1}^{(1)},\\
 \label{same-3}
\Psi_{n+1}^{(1)}&=&\left[E_{n}^{(2)}\right]^{-\frac{1}{2}}A^\dag\Psi_n^{(2)}.
\end{eqnarray}
In what follows, let us express some facts.
(i) If the ground-state wave function $\Psi_0^{(1)}$, which is given by
($A\Psi_0^{(1)}=0$)
\begin{eqnarray}\label{psi-0}
 \Psi_0^{(1)}=N_0 {\rm exp}\left[-\int^x W(x')dx'\right],
\end{eqnarray}
is square integrable, then the ground state of $H_1$ has zero energy ($E_0 =0$) \cite{CGK87}.
For this case, it can be shown that the SUSY is unbroken.
(ii) if the eigenfunction $\Psi_{n+1}^{(1)}$ of $H_1$ ($\Psi_n^{(2)}$ of $H_2$)
is normalized, then the $\Psi_n^{(2)}$ ($\Psi_{n+1}^{(1)}$), will be also normalized.
(iii) Assuming the eigenfunction $\Psi_n^{(1)}$ with eigenvalue $E_n^{(1)}$ ($\Psi_n^{(2)}$ with eigenvalue $E_n^{(2)}$)
corresponds to the {Hamiltonian} $H_1$ ($H_2$), it is easy to show
that $A\Psi_{n}^{(1)}$ ($A^\dag\Psi_{n}^{(2)}$) will be an eigenfunction of $H_2$ ($H_1$) with the same eigenvalue.
(iv) In order to destroy (create) an extra node in the eigenfunction as well as convert
an eigenfunction of $H_1$ ($H_2$) into an eigenfunction of $H_2$ ($H_1$) with the same
energy, we apply the operator $A$ ($A^\dag$).
(v) The ground state wave function of $H_1$ has no SUSY partner. (v) Applying
the operator $A$ ($A^\dag$), all the eigenfunctions of $H_2$ ($H_1$, except for the
 ground state) can be reconstructed from those of $H_1$ ($H_2$); {please} see Fig.(\ref{fig1}).

 It has been believed that {this fascinating procedure, which
  leads to understand 
  the degeneracy} of the
 spectra of $H_1$ and $H_2$, can be provided from applying the properties of the SUSY algebra.
 Therefore, let us consider a matrix SUSY Hamiltonian (which is part of a closed
 algebra including both bosonic and fermionic operators with
 commutation and anti-commutation relations) 
 {containing} both the Hamiltonians $H_1$ and $H_2$ \cite{CKS95}:
\begin{eqnarray}\label{H-matrix}
H=\left[
  \begin{array}{cc}
    H_1 & 0 \\
    0 & H_2 \\
  \end{array}
\right].
\end{eqnarray}
Supersymmetric quantum mechanics begins with a set of two matrix, $Q$ and $Q^\dagger$, operators,
known as supercharges
\begin{eqnarray}\label{Q}
Q=\left[
  \begin{array}{cc}
    0 & 0 \\
    A & 0 \\
  \end{array}
\right],
\end{eqnarray}
\begin{eqnarray}\label{Q-dag}
Q^\dag=\left[
  \begin{array}{cc}
    0 & A^\dag \\
    0 & 0 \\
  \end{array}
\right].
\end{eqnarray}
The matrix $H$ is part of a closed algebra in which both bosonic and fermionic operators with commutation
and anti-commutation relations {are included, such that the bosonic degrees of freedom are changed into the
fermionic ones and vice versa by the supercharges.}

It is straightforward to show that
\begin{eqnarray}\label{com-anti-1}
[H,Q]\!\!\!&=&\!\!\![H,Q^\dag]=0,\\
\label{com-anti-2}
\{Q,Q^\dag\}\!\!\!&=&\!\!\!H,\\
%\end{eqnarray}
%\begin{eqnarray}
\label{com-anti-3}
\{Q,Q\}\!\!\!&=&\!\!\!2Q^2=\{Q^\dag,Q^\dag\}=2\left(Q^\dag\right)^2=0,
\end{eqnarray}
by which the closed superalgebra $sl(1,1)$ is described \cite{K77} (see also subsection \ref{SUSY-QC}).
Note that the relations \eqref{com-anti-1} are responsible for the degeneracy.

In {SQM}, when the two partner potentials have continuum
spectra, it is possible to relate the reflection and transmission coefficients.
A necessary condition for providing scattering in both
of the partner potentials is that they must be finite
when $x\rightarrow -\infty$ or $x\rightarrow \infty$.

\subsection{Shape invariance and solvable potentials}
\label{SIP}

\indent

In the context of the non-relativistic quantum
mechanics, there are a number of known potentials (e.g., Coulomb,
harmonic oscillator, Eckart, Morse, and P\"{o}schl--Teller) for which we can solve the corresponding Schr\"{o}dinger
equation analytically and determine all the energy eigenvalues and eigenfunctions explicitly. In this regard, the following questions naturally arise: Why are just some potentials solvable? Is there any underlying symmetry property? What is this symmetry?

%In the context of the non-relativistic quantum mechanics, there are a number of known potentials (e.g., Coulomb, harmonic oscillator, Eckart, Morse and P\"{o}schl--Teller) for which we can solve analytically the corresponding Schr\"{o}dinger equation and determine explicitly all the energy eigenvalues and eigenfunctions. In this regard, the following questions naturally arise: Why just some potentials are solvable? Is there any underlying symmetry property? What is this symmetry?

 Gendenshtein was the first to answer these questions by
 introducing the {\it shape invariance} concept \cite{G83}.
 In fact, for such potentials, all the bound state energy eigenvalues, eigenfunctions and the scattering
 matrix can be retrieved from applying the generalized operator
 method, which is essentially equivalent to the Schr\"{o}dinger's method of
factorization \cite{S40,IH51}.

In \cite{G83}, the relationship between SUSY, the
hierarchy of Hamiltonians, and solvable potentials
has been investigated from an interesting perspective
(for detailed discussions see, for instance, \cite{CGK87,CKS95}).
In what follows, let us describe briefly the shape invariance concept.
``{\it If the pair of SUSY
partner potentials $V_{1,2}(x,b)$ are similar in shape and differ only in the
parameters that appear in them, then they are said to be shape invariant}'' \cite{K97}.
Let us be more precise. 
 {Consider a pair of SUSY partner potentials, $V_{1,2}(x)$, as defined in \eqref{V-vs-W}. If the profiles of these potentials are such that they satisfy the relationship
\begin{eqnarray}\label{SIC}
V_2(x,b)=V_1(x,b_1)+R(b_1),
\end{eqnarray}
where the parameter $b_1$ is some function of $b$, say given by $b_1 =
f(b)$, the potentials $V_{1,2}(x)$ are said to bear shape invariance. In other words, to be associated within shape invariance
the potentials $V_{1,2}$,  while sharing a similar coordinate dependence,  can
at most differ in the presence of some parameters. To make the definition of shape invariance clear, consider for example
\begin{equation}\label{sha1}
    W=b\tanh\left(\frac{\sqrt{2m}}{\hbar} x\right).
\end{equation}
Then, inserting this superpotential into \eqref{V-vs-W} gives us
\begin{equation}\label{sha2}
  \begin{split}
      V_1(x,b)=-\frac{b(b+1)}{\cosh^2\left(\frac{\sqrt{2m}}{\hbar} x\right)}+b^2,\\
       V_2(x,b)=-\frac{b(b-1)}{\cosh^2\left(\frac{\sqrt{2m}}{\hbar} x\right)}+b^2
  \end{split}  
\end{equation}
The above be expressions show that one can rewrite 
$V_2$ in therms of $V_1$, as expressed in \eqref{V-vs-W}
where, in this example, $b_1=b-1$, and $R(b_1):=2b_1+1$. Thus, the potentials $V_{1,2}$ bear shape invariance in accordance with the definition \eqref{V-vs-W}. Then, to use the shape invariance condition, let us assume that \eqref{V-vs-W} holds for a
sequence of parameters, $\{b_k\}_{k = 0, 1, 2, . . .}$, where 
\begin{equation}
   b_k= \underbrace{f\circ f\circ f\circ...\circ f}_{\text{k times}}(b)=f^k(b),~~~k=0,1,2,...,~~~b_0:=b.
\end{equation}
Consequently, 
\begin{equation}
    \label{sha4}
    H_2(x,b_k)=H_1(x,b_{k+1})+R(b_k).
\end{equation}
Now, we write $H^{(0)}=H_1(x,b)$, $H^{(1)}=H_2(x,b)$, and we define $H^{(m)}$ as
\begin{equation}\label{sha5}
H^{(m)}:=-\frac{\hbar^2}{2m}\frac{d^2}{dx^2}+V_1(x,b_m)+\sum_{k=1}^mR(b_k)=H_1(x,b_m)+\sum_{k=1}^mR(b_k).
\end{equation}
Using (\ref{sha4}), we can extract $H^{(m+1)}$ as
\begin{equation}
    \label{sha6}
    H^{(m+1)}=H_2(x,b_m)+\sum_{k=1}^mR(b_k).
\end{equation}
Therefore, in this way, we are able to set up a hierarchy of Hamiltonians $H^{(k)}$ 
for various $k$ values.
}
%If the requirement
%\begin{eqnarray}\label{SIC}
%V_2(x,b)=V_1(x,b_1)+R(b_1),
%\end{eqnarray}
%(where $b$ is a set of parameters, $b_1$ is a function of $b$ (say $b_1 = f(b)$) and the remainder $R(b_1)$ is a constant independent of $x$) is satisfied by $V_1(x,b)$ and $V_2(x,b)$, then they are called shape invariant potentials.

Employing condition \eqref{SIC} and the hierarchy of Hamiltonians \cite{CKS95},
the energy eigenvalues and eigenfunctions have been obtained for any shape invariant potential
when SUSY is unbroken.
It should be noted that there is a correspondence
between the condition \eqref{SIC} (associated with 
SQM) and the required mathematical
 condition applied in the method of the factorization of the Hamiltonian \cite{S41}.
Although the terminology and ideas associated with these methods {are
different, 
they can} be considered as the special cases of the
procedure employed to handle second-order linear differential equations \cite{D1882,LP86}.
Notwithstanding the above, it has been believed that 
{a better} understanding
of analytically solvable potentials could be achieved by SUSY and shape invariance. {Let us elaborate more on this aspect.}

 {$H_2$ contains the lowest state with a zero energy eigenvalue, according to the SQM concepts discussed in  subsection (\ref{SUSY-QM}). As a result of (\ref{same-1}), the lowest energy level of $H^{(m)}$ has the value of
\begin{equation}
    \label{sha7}
  E_0^{(m)}=\sum_{k=1}^mR(b_k).  
\end{equation}
Therefore, it is simple to realize that because of the chain
$H^{(m)}\rightarrow H^{(m-1)}...\rightarrow H^{(1)}(:= H_2)\rightarrow H^{(0)}(:= H_1)$, the $n$th member in this sequence carries the nth level of the energy spectra of $H^{(0)}$
(or $H_1$), namely \cite{CGK87}
%Therefore, employing \eqref{same-1}, a straightforward consequence of \eqref{SIC} is
%that the complete bound-state spectrum of $H_1$ (or $H^{(0)}$) is given by \cite{CGK87}
\begin{eqnarray}\label{SIC-2}
E_{n}^{(0)}=\sum_{k=1}^{n}R(b_k),~~~~E_0^{(0)}=0.
\end{eqnarray}
Let us now return to the example (\ref{sha1}). We rewrite (\ref{SIC}) as 
\begin{equation}
    V_1(x,b)=V_2(x,b-1)+b^2-(b-1)^2.
\end{equation}
We can generate $b_k$ from $b_0=b$ as $b_k=b-k$. Hence, the energy spectrum from $V_1(x,b)$ yields
\begin{equation}
    E^{(0)}_n=\sum_{k=1}^nR(b_k)=\sum_{k=1}^n(b^2-b_k^2)=b^2-b_n^2=b^2-(b-n)^2.
\end{equation}
}
It is worthy to note that, according to the requirement \eqref{SIC}, the well-known solvable
potentials (such as those were listed in the first paragraph of this subsection)
are all shape invariant, and, therefore, their
energy eigenvalue spectra is given by \eqref{SIC-2}.
``{\it In \cite{G83},
Gendenshtein then conjectured that shape invariance
is not only sufficient but may even be necessary for
a potential to be solvable} \cite{CGK87}.''
Moreover, by applying SUSY, it is also possible to retrieve
 the bound-state energy eigenfunctions of $H_1$ for shape invariant
potentials \cite{DKS88}.
In particular, in the same paper, by taking $\Psi_0^{1}(x,b)$ as
the ground-state wave function of
$H_1$ (which is given by \eqref{psi-0}), and employing
relation \eqref{same-3}, a relation is obtained for $n$th-state eigenfunction $\Psi_n^{1}(x,b)$ as
\begin{eqnarray}\label{SIC-3}
\Psi_n^{1}(x,b)=A^\dag (x,b)A^\dag (x,b_1)...A^\dag (x,b_{n-1})\Psi_0^{1}(x,b).
\end{eqnarray}

For  {later} convenience, let us concentrate on a specific shape invariance that only
involving translation of the parameter $b_0$ with 
{a} translation step $\eta$ \cite{FR01}
(for other kind of relations between the parameters, see, for instance, \cite{CKS95}):
\begin{eqnarray}\label{trans-SI}
b_1=b_0+\eta.
\end{eqnarray}
It is feasible to introduce a translation operator as
\begin{eqnarray}\label{trans-op}
T(b_0)={\rm exp}\left(\eta \frac{\partial}{\partial b_0}\right), \hspace{10mm}
T^{-1}(b_0)=T^{\dag}(b_0)={\rm exp}\left(-\eta \frac{\partial}{\partial b_0}\right),
\end{eqnarray}
which act merely on objects defined on the parameter space.

By composing the translation and bosonic operators, we can
construct the {following} generalized creation and annihilation operators:
\begin{eqnarray}\label{gen-op-1}
B_1(b_0)=A^{\dag}(b_0)T(b_0),\cr\\
\label{gen-op-2}
B_2(b_0)=T^{\dag}(b_0)A(b_0).
\end{eqnarray}
Applying the shape invariant potentials to solve the Schr\"{o}dinger
equation is similar to the factorization method employed to the
 case of the harmonic oscillator potential \cite{FR01}. Therefore, we have
\begin{eqnarray}\label{gen-op-3}
B_2(b_0)A(b_0)\Psi_0(x;b_0)=A(b_0)\Psi_0(x;b_0)=0.
\end{eqnarray}
In order to obtain the excited states, the creation operator should repeatedly act on $\Psi_0(x;b_0)$:
\begin{eqnarray}\label{gen-op-4}
\Psi_n(x;b_0)=\left[B_1(b_0)\right]^n\Psi_0(x;b_0).
\end{eqnarray}
We should note that the translation operators ($T$ and $T^{\dag}$) and ladder operators, $A$ and $A^\dagger$ do not
commute with any {$b_k$-dependent} and any $x$-dependent objects, respectively.
Therefore, the generalized creation and annihilation operators ($B_1$ and $B_2$) act on the
objects defined on the dynamical variable space and the objects defined on parameter
space via the bosonic operators and the translation operators, respectively.
According to \eqref{gen-op-3}, we have
\begin{eqnarray}\label{gen-op-4a}
\Psi_0(x;b_0)\varpropto {\rm exp}\left(-\int^xW(\tilde{x};b_0)d\tilde{x}\right),
\end{eqnarray}
which is transformed by a
normalization constant (that should, in general, depend on parameters $b$)
 into a relation of equality. Concretely, the action of the generalized
 operators affects in determining such a normalization constant; for more details, see, \cite{FA93}.

Now let us obtain the relations of the energy eigenvalues and energy spectrum.
From using \eqref{trans-op}, we can write
\begin{eqnarray}\label{gen-op-5}
R(b_n)=T(b_0)R(b_{n-1})T^{\dag}(b_0),
\end{eqnarray}
where
\begin{eqnarray}\label{trans-SI-2}
b_n=b_{0}+n\eta
\end{eqnarray}
is a generalized version of \eqref{trans-SI}.
Employing \eqref{gen-op-5}, we get
\begin{eqnarray}\label{gen-op-6a}
R(b_n)B_1(b_0)=B_1(b_0)R(b_{n-1}).
\end{eqnarray}

Equations \eqref{gen-op-5} and \eqref{gen-op-6a} yield a commutation relation as
\begin{eqnarray}\label{gen-op-6}
[H_1,\left(B_1\right)^n]=\left(\sum_{k=1}^{n}R(b_k)\right)\left(B_1\right)^n.
\end{eqnarray}
Applying \eqref{gen-op-6} on the ground state of $H_1$, i.e., 
$\Psi_0(x;b_0)$, it is seen that $\left[B_1(b_0)\right]^n \Psi_0(x;b_0)$ is also an 
eigenfunction of $H_1$ with eigenvalue $E_{n}^{(1)}$ given by \eqref{SIC-2}. 
Therefore, the energy spectrum is:
\begin{eqnarray}\label{gen-E-1}
E_n=E_0+E_{n}^{(1)},
\end{eqnarray}
where the ground state energy $E_0$ is obtained from either
\begin{eqnarray}\label{gen-E-2}
H_1=H-E_0,
\end{eqnarray}
or, equivalently,
\begin{eqnarray}\label{gen-E-3}
W(x;b)-W'(x;b)=V(x)-E_0=V_1(x).
\end{eqnarray}

Finally, it should be noted that the above established algebraic approach is 
self-consistent. {More concretely,}  by considering supersymmetric and shape invariance properties of the system,
it can be applied as an appropriate method for obtaining not only the 
eigenvalues and eigenfunctions of the bound state 
of a Schr\"{o}dinger equation, but also exact resolutions for this equation
\cite{FR01}.

\section{SUSY Quantum Cosmology}
\label{QC}

\indent

In order to apply the formalism presented in the previous section, let us 
investigate a homogeneous and isotropic cosmology, in the context of {General Relativity (GR)} together with a single
 scalar field, $\phi$, minimally coupled to gravity.  
 
 \subsection{A case study: classical setting}
 \label{QC-1}

\indent
 
 By considering the Friedmann–Lema\^{i}tre–Robertson–Walker (FLRW) line 
 element\footnote{Throughout this paper we work in natural units where $\hbar=c = k_B = 1$.} 
 \begin{equation}
     ds^2=N(t)dt^2+a(t)^2\left\{ \frac{dr^2}{1-kr^2}+r^2 d\Omega^2 \right\},
 \end{equation}
 the {ADM}\footnote{Adler-Deser-Misner (ADM); see \cite{doi:10.1142/8540}  for more details.} Lagrangian will be \footnote{In this work, we consider a framework in which the scalar field is minimally coupled to gravity, see also \cite{RSFMM18,RPSM20,R22}. Instead, one can choose other interesting gravitational models where $\phi$ is non-minimally coupled, see for instance, \cite{RFK11, RM14, RZJM16, RM16, RMM19}.} %\cite{doi:10.1142/8540}
\begin{eqnarray}\label{ADM-FRW}
L_{\rm ADM}=-\frac{3}{N}a\dot{a}^2+3kNa+a^3\left(\frac{\dot{\phi}^2}{2N}-NV(\phi)\right),
\end{eqnarray}
where an over-dot denotes a differentiation with respect to the
 cosmic time $t$; $N(t)$ is a lapse function, $a(t)$ is the scale
factor, $V(\phi)$ is a scalar potential and $k=\{-1,0,1\}$ is the spatial curvature constant
associated with open, flat and closed universes, respectively.

In this work, let us consider the scalar potential $V(\phi)$ to be in the form \cite{DT93,DOT93}
\begin{eqnarray}\label{DOT-pot}
V(\phi)=\lambda+\frac{m^2}{2\alpha^2}\sinh^2(\alpha\phi)+\frac{\vartheta}{2\alpha^2}\sinh(2\alpha\phi),
\end{eqnarray}
where $\lambda$ may be related to the cosmological constant;
$m^2=\partial^2V/\partial\phi^2|_{\phi=0}$ is a mass squared
 parameter; $\alpha^2=3/8$ and $\vartheta$ is a coupling parameter.
 Moreover, we {will now} investigate only the spatially flat FLRW
 universe. For this case, it has been shown that an oscillator--ghost--oscillator system is produced \cite{DOT93,Jalalzadeh:2003iu,Khosravi:2007vh,Pedram:2007mj,Bina:2007wj,Jalalzadeh:2011yp}. { More concretely, by applying the following transformations \cite{Khosravi:2006es,Vakili:2010qf,Darabi:1999yt,Darabi:2004bh}
 \begin{eqnarray}\label{DOT-transf}
X=\frac{a^{\frac{3}{2}}}{\alpha}\cosh(\alpha\phi),\hspace{10mm}Y=\frac{a^{\frac{3}{2}}}{\alpha}\sinh(\alpha\phi),
\end{eqnarray}
the Lagrangian \eqref{ADM-FRW} transform into
\begin{equation}
    \label{shahram1}
L_{\rm ADM}=-\frac{1}{2N}\dot\xi^{\top}J\dot\xi+\frac{N}{2}\xi^\top MJ\xi,    
\end{equation}
where
\begin{eqnarray}
 \label{shahram2}
\xi:=\begin{pmatrix}
X\\ Y
\end{pmatrix},~~~~~
M:=\begin{pmatrix}
 2\lambda\alpha^2 & -\vartheta\\
 \vartheta & 2\lambda\alpha^2-m^2
 \end{pmatrix},~~~~~J:=\begin{pmatrix}
 1&0\\
 0&-1
 \end{pmatrix}.
\end{eqnarray}
It is straightforward to decouple (\ref{shahram1}) into normal modes $\gamma:=\Sigma^{-1}\xi$ by means of
\begin{equation}
    \label{shahram3}
  \gamma:=\begin{pmatrix}
u\\v
\end{pmatrix},~~~~~\Sigma:=\begin{pmatrix}
\frac{-m-\sqrt{m^4-4\vartheta^2}}{2\vartheta}& \frac{-m+\sqrt{m^4-4\vartheta^2}}{2\vartheta}\\
1&1
\end{pmatrix},  
\end{equation}
which diagonalize the matrix $M$ as follows
\begin{equation}
    \label{shahram4}
    \Sigma^{-1}M\Sigma=\begin{pmatrix}
    \omega_1&0\\
    0&\omega_2
    \end{pmatrix},~~~~\omega_{1,2}^2=\frac{3\lambda}{4}+\frac{m^2}{2}\mp\frac{\sqrt{m^4-4\vartheta^2}}{2}.
\end{equation}
}

{Thus, we retrieve the Lagrangian associated with a $2D$ oscillator--ghost--oscillator:
\begin{equation}\label{ADM-FRW-2}
\begin{split}
L_{\rm ADM}(u,v)&=-\frac{1}{2N}\dot\gamma^\top\mathcal I\dot\gamma+\frac{N}{2}\gamma^\top\mathcal J\gamma\\
&=-\frac{1}{2}\left\{\left(\frac{1}{N}\dot{u}^2-\omega_1^2Nu^2\right)-\left(\frac{1}{N}\dot{v}^2-\omega_2^2Nv^2\right)\right\},
\end{split}
\end{equation}
where $\mathcal I:=\Sigma^\top J\Sigma$, and $\mathcal J:=\Sigma^\top MJ\Sigma$.
}
%where $\omega_{1,2}^2=3\lambda/4+\frac{m^2}{2}\mp\sqrt{m^4-4b^2}/2$. 
The conjugate momenta corresponding to $u$ and $v$ are:
\begin{eqnarray}\label{momenta}
p_u=\frac{\dot{u}}{N},\hspace{10mm} p_v=-\frac{\dot{v}}{N}.
\end{eqnarray}
Moreover, the classical Euler--Lagrange equations are given by
\begin{eqnarray}\label{class-EL-1}
\frac{d}{dt}\left(\frac{\dot{u}}{N}\right)+N\omega_{1}^2 u=0,~~~~~
\label{class-EL-2}
\frac{d}{dt}\left(\frac{\dot{v}}{N}\right)+N\omega_{2}^2 v=0.
\end{eqnarray}

It is straightforward to show that the Hamiltonian corresponding to the ADM Lagrangian \eqref{ADM-FRW-2} is: 
\begin{eqnarray}\label{class-Ham}
H_\text{ADM}=-\frac{N}{2}\left\{\left(p_u^2+\omega_1^2u^2\right)-\left(p_v^2+\omega_2^2v^2\right)\right\},
\end{eqnarray}
which, for the gauge $N=1$, yields
\begin{eqnarray}\label{class-sol}
u(t)=u_0\sin(\omega_{1}t-\theta), \hspace{10mm}v(t)=v_0\sin(\omega_{2}t).
\end{eqnarray}
In \eqref{class-sol}, $\theta$ is an arbitrary phase factor. From using the Hamiltonian constraint, we obtain $\omega_{1}u_0=\omega_{2}v_0$. 
It is also seen that the classical paths corresponding to
solutions \eqref{class-sol}, in the configuration
space $(u,v)$, are the generalized Lissajous ellipsis.

 \subsection{Quantization}
 \label{QC-2}

\indent

In order to establish a quantum cosmological
model corresponding to our herein model, let us proceed with the Wheeler--DeWitt equation. The canonical quantization of (\ref{class-Ham}) gives
\begin{eqnarray}\label{WD}
{\cal H}\Psi(u,v)=\left(-\frac{\partial^2}{\partial u^2}+\frac{\partial^2}{\partial v^2}
+\omega_{1}^2 u^2-\omega_{2}^2 v^2\right)\Psi(u,v)=0.
\end{eqnarray}
Equation \eqref{WD} is separable and we can obtain a solution as
\begin{eqnarray}\label{WD-sol}
\Psi_{n_1,n_2}(u,v)=\alpha_{n_1}(u)\beta_{n_2}(v),
\end{eqnarray}
where
\begin{eqnarray}\label{alpha}
\alpha_{n}(u)&=&\left(\frac{\omega_1}{\pi}\right)^{1/4}\frac{H_n(\sqrt{\omega_{1}}u)}{\sqrt{2^n n!}}{\rm e}^{-\omega_{1} u^2/2},\\\nonumber\\
\label{beta}
\beta_{n}(v)&=&\left(\frac{\omega_2}{\pi}\right)^{1/4}\frac{H_n(\sqrt{\omega_{2}}v)}{\sqrt{2^n n!}}{\rm e}^{-\omega_{2} v^2/2}.
\end{eqnarray}
In relations \eqref{alpha} and \eqref{beta}, $H_n(x)$ stands for the Hermite polynomials.
Moreover, we should note that the Hamiltonian constrain relates the parameters of the model as
\begin{eqnarray}\label{H-C}
\left(n_1+\frac{1}{2}\right)\omega_{1}=\left(n_2+\frac{1}{2}\right)\omega_{2}, \hspace{10mm} n_1,n_2=0,1,2, ...~  .
\end{eqnarray}
{The recovery of classical solutions from the corresponding quantum model is one of the essential elements of quantum cosmology. For this aim,  a coherent wave packet with reasonable asymptotic behavior in the minisuperspace is often constructed, peaking near the classical trajectory. We  can herewith produce a widespread wave packet solution
\begin{equation}
    \label{sha8}
    \Psi(u,v)=\sum_{n_1,n_2}C_{n_1n_2}\alpha_{n_1}(u)\beta_{n_2}(v),
\end{equation}
where the summing is restricted to overall values of $n_1$ and
$n_2$ satisfying the relation (\ref{H-C}). Let us consider the simplest case which is when $\omega_1=\omega_2=\omega$, which means $m^2=2\vartheta$ in definition of scalar field potential (\ref{DOT-pot}). Then, the wave packet will be
\begin{equation}
    \label{sha9}
    \Psi(u,v)=\sqrt{\frac{\omega}{\pi}}\sum_{n=0}^{\infty}\frac{C_n}{n!2^n}\exp{\left(-\frac{\omega}{2}(u^2+v^2)\right)}H_n(\sqrt{\omega}u)H_n(\sqrt{\omega}v),
\end{equation}
where $C_n$ is a complex constant.  We apply the following identity to create a coherent wave packet with suitable asymptotic behavior in the minisuperspace, peaking around the classical trajectory
\begin{equation}
    \label{sha10}
 \sum_{n=0}^{\infty}\frac{t^n}{n!}H_n(x)H_n(y)=   \frac{1}{\sqrt{1-t^2}}\exp{\left(\frac{2txy-t^2(x^2+y^2)}{2(1-t^2)}\right)}.
\end{equation}
Using this identity and choosing the coefficients $C_n$ in (\ref{sha9}) to be
$C_n= B2^n\tanh\xi$ , with $B$ and $\xi$ are arbitrary complex constants, we obtain
\begin{multline}
    \label{sha11}
    \Psi(u,v)=C\exp{\left(-\frac{\omega}{4}\cos(2\beta_2)\cosh(2\beta_1)(u^2 + v^2-2\eta\tanh(2\beta_1)uv)\right)}\\
\times\exp{\left( -\frac{i\omega}{4}\sinh(2\beta_1) \sin(2\beta-2)(u^2 + v^2 - 2\eta \coth(2\beta_1)uv)\right)},
\end{multline}
where $\beta_1$ and $\beta_2$ are the real and imaginary parts of $\xi=\beta_1+i\beta_2$, respectively, $\eta=\pm1$, and  $N$ is a normalization factor. Fig.(2.a) shows the density plot, and Fig.(2.b) illustrates the contour plot of the wave function  for typical values of $\beta_1$, $\beta_2$, and $\eta=1$ for the following combination of the solutions
\begin{equation}
    \Psi(u,v)=\Psi_{\beta_1,\beta_2}(u,v)-\Psi_{\beta_1+\delta\beta_1,\beta_2+\delta\beta_2}(u,v).
\end{equation}

\begin{figure}
\subfloat[]{\includegraphics[width = 2.5in]{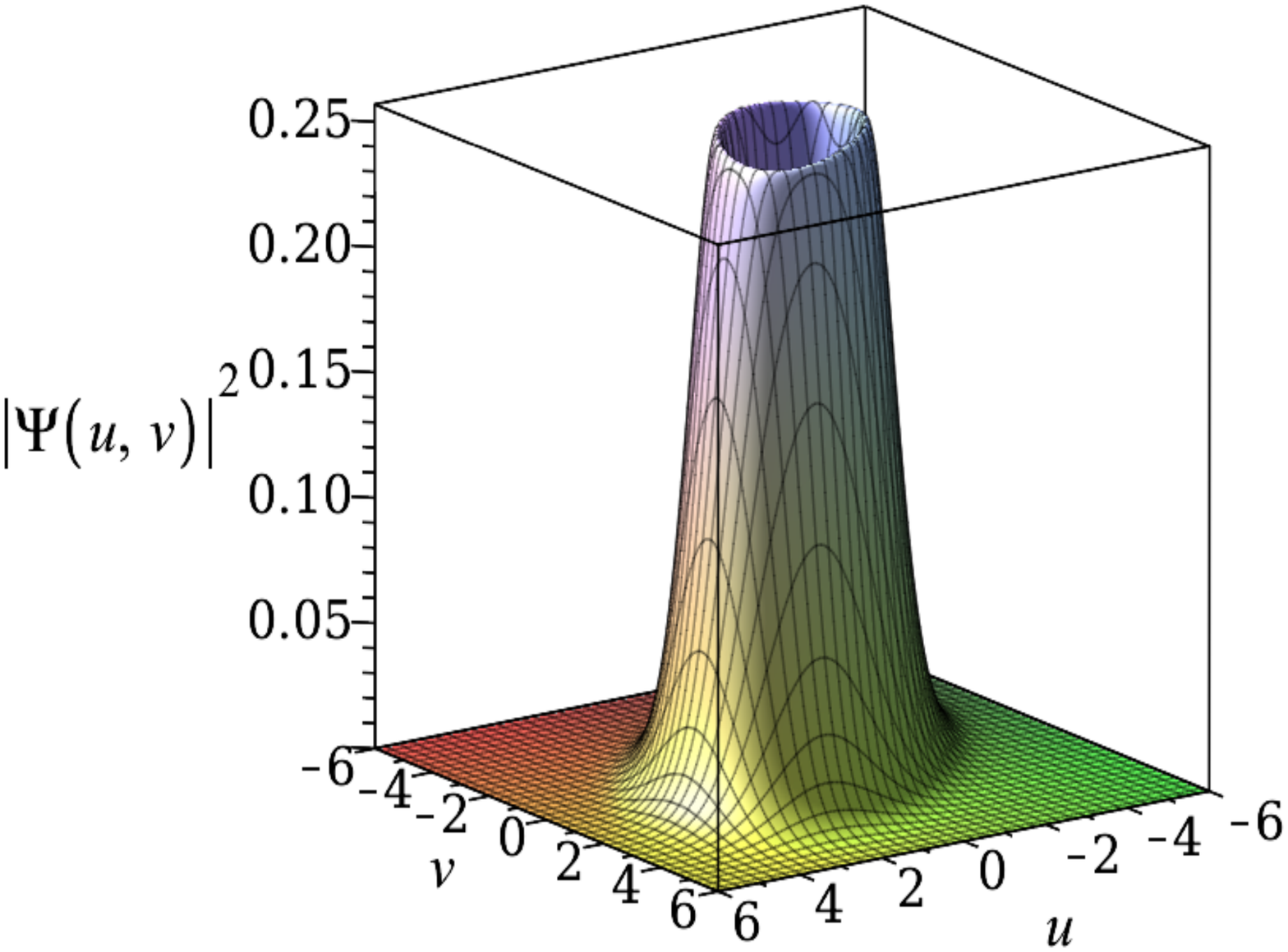}} 
\subfloat[]{\includegraphics[width = 2.3in]{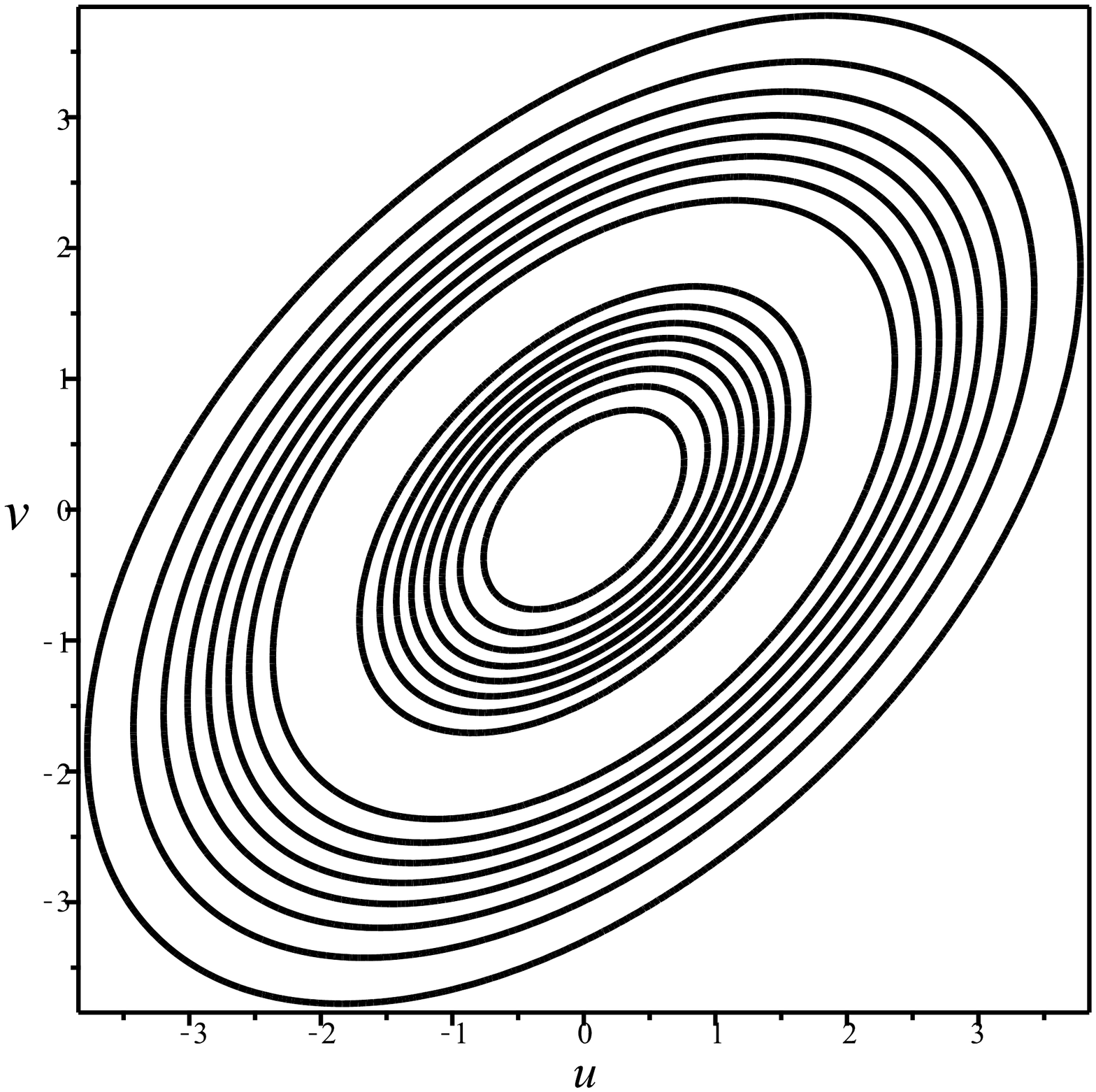}}\\
\caption{Density plot, Fig.(a), and contour plot, Fig.(b), of a wave packet. These figures are plotted for numerical values $\omega=1$, $\beta_1=1$, $\beta_2= \pi/6$, $\eta=1$, $\delta\beta_1=0.1$ and $\delta\beta_2=3\pi/50$.}
\label{fig2}
\end{figure}

The classical solutions (\ref{class-sol}) can easily be represented as the following trajectories (for $\omega_1=\omega_2=\omega$)
\begin{equation}\label{sha13}
    u^2+v^2-2\eta uv\cos(\theta)-u_0^2\sin^2(\theta)=0.
\end{equation}
This equation describes ellipses whose major axes make angle $\pi/4$ with the positive/negative $u$ axis according to the choices $\pm1$ for $\eta$. Also, each trajectory's eccentricity and size are determined by $\theta$ and $u_0$, respectively. It can be seen that the quantum pattern in Fig.(\ref{fig2}) and the classical paths (\ref{sha13}) in configuration space, $(u,v)$, have a high correlation.}

Let us point  how we can introduce a time-evolving wave-function.  By employing a canonical transformation on the $(v,p_v)$ sector of the Hamiltonian \eqref{class-Ham}, we observe that in the total Hamiltonian, the momentum associated with the  the new canonical variable appears linearly. Let us be more precise.  {Consider the following canonical transformation $(v,p_v)\rightarrow(T,p_T)$ given by 
\begin{equation}\label{transf}
 v=\sqrt{\frac{2p_T}{\omega_2^2}}\sin(\omega_{2}T),~~~~~
  p_v=\sqrt{2p_T}\cos(\omega_{2}T),~~~~\{T,p_T\}=1.
 \end{equation}
 It is easy to check out that the inverse map is given by following relations
 \begin{equation}
     p_T=\frac{1}{2}p_v^2+\frac{1}{2}\omega_1^2v^2,~~~T=\frac{1}{\omega_2}\tan^{-1}\left(\omega_2\frac{v}{p_v}\right).
 \end{equation}
 In fact, the new set of phase space coordinates $(T, p_T)$ is related to the harmonic oscillator's action-angle variables, $(\varphi, p_\varphi)$, by \cite{JRA20,2007AnPK}
 \begin{equation}
    \varphi=\omega_2T,~~~~p_\varphi=\frac{p_T}{\omega_2}.
 \end{equation}
 }
The ADM Hamiltonian \eqref{class-Ham} {simply} takes the form
\begin{eqnarray}\label{T-Ham}
H_\text{ADM}=N\left(\frac{1}{2}p_u^2+\frac{1}{2}\omega_1^2u^2-p_{_{T}}\right).
\end{eqnarray}
The classical field equations corresponding to \eqref{T-Ham} are
\begin{eqnarray}\label{transf1}
\begin{cases}
  \dot{u}=N p_u,~~~~
  \dot{p}_u=-N\omega_1^2u,\\
 \dot{T}=-N,~~~~
 \dot{p}_{_{T}}=0.
 \end{cases}
\end{eqnarray}
For $N=1$ we find 
\begin{eqnarray}\label{T-sol-1}
T&=&-t,~~~~p_{_{T}}={\rm const.}
\end{eqnarray}
%The first set of solutions of \eqref{transf} are the same solutions obtained before.
{Thus, the motion in $2D$ phase space $(T,p_T)$ becomes trivial, i.e.,flow paths are
straight lines with constant $p_T$.}
As seen, the second set of solutions for \eqref{transf1} implies that $T$ plays the role of the time parameter.
Consequently, the Poisson bracket of the
time parameter and super-Hamiltonian does not vanish but instead we
have $\{T,{\cal H} \}=1=\{T,p_{_{T}} \}$, which implies that
$T$ is not a Dirac observable, and therefore, we may consider it as a time
{variable; see for instance, \cite{JRA20} and references therein}.

\subsection{{Supersymmetric quantization}}
\label{SUSY-QC}

\indent 

Employing the Hamiltonian constraint upon (\ref{T-Ham}), and then substituting $p_u=-i\frac{
d}{du}$ and $p_T=-i\frac{\partial}{\partial T}=i\frac{\partial}{\partial t}$, we get a Schr\"odinger--Wheeler--WeDitt equation:
\begin{equation}\label{sh0}
    i\frac{\partial}{\partial t}\Psi(u,t)=\left[-\frac{1}{2}\frac{d^2}{du^2}+\frac{l(l+1)}{2u^2}+\frac{1}{2}\omega_1^2u^2  \right]\Psi(u,t).
\end{equation}
%Let us start with the Hamiltonian \eqref{T-Ham}, for which the Hamiltonian
%constraint is satisfied. Therefore, we obtain the Hamiltonian ${\cal H}_1$ as
%\begin{eqnarray}\label{new-Ham}
%2p_{_{T}}=\omega_1\left(p^2+x^2\right)\equiv{\cal H}_1,
%\end{eqnarray}
%where, instead of $u$ and $p_u$, we have used their corresponding dimensionless variables:
%\begin{eqnarray}\label{dLess}
%p=\frac{p_u}{\sqrt{\omega_1}}, \hspace{10mm} x=\sqrt{\omega_1}u.
%\end{eqnarray}
In the process of obtaining equation \eqref{sh0}, we have further used the following factor ordering procedure:
 \begin{eqnarray}\label{Fac-Ord}
p_u^2=-\frac{1}{3}\left(u^\alpha\frac{d}{du}u^\beta\frac{d}{du}u^\gamma+ u^\gamma\frac{d}{du}u^\alpha\frac{d}{du}x^\beta+u^\beta\frac{d}{du}u^\gamma\frac{d}{du}u^\alpha\right),
\end{eqnarray}
where the parameters $\alpha$, $\beta$, and $\gamma$ satisfy the requirement $\alpha+\beta+\gamma=0$, and we have set  $\frac{1}{3}(\beta^2+\gamma^2+\beta\gamma):=l(l+1)$.
%the Hamiltonian \eqref{new-Ham} is modified as
%\begin{eqnarray}\label{new-Ham-1}
%{ H}=\left[-\frac{d^2}{dx^2}+x^2+\frac{l(l+1)}{x^2}\right],~~~~~l(l+1):=\frac{1}{3}(\beta^2+\gamma^2+\beta\gamma),
%\end{eqnarray}
%where for convenience, we introduced a new Hamiltonian as ${H}\equiv{\cal H}_1/\omega_1$, which corresponds to the radial harmonic oscillator and can be easily analyzed
%using the presented procedures associated with the shape invariant potential and
%generalized ladder operators, see for more detail \cite{FR01}.

The time independent sector of equation \eqref{sh0} reads
\begin{equation}\label{sh2}
    H_l\Psi^l_n(u)=E_n^l\Psi^l_n(u),
\end{equation}
where
\begin{equation}\label{sham}
    H_l:=-\frac{1}{2}\frac{d^2}{du^2}+\frac{l(l+1)}{2u^2}+\frac{1}{2}\omega_1^2u^2.
\end{equation}
According to the Eq.(\ref{V-vs-W}), we {now} introduce first-order differential operators
\begin{equation}\label{sh1}
\begin{cases}
  A_l:=\frac{1}{\sqrt{2\omega_1}}\frac{d}{du}+\sqrt{\frac{\omega_1}{2}}u-\frac{l+1}{\sqrt{2\omega_1}u}\,\,,\\\\
  A^\dagger_l:=-\frac{1}{\sqrt{2\omega_1}}\frac{d}{du}+\sqrt{\frac{\omega_1}{2}}u-\frac{l+1}{\sqrt{2\omega_1}u}\,\,.
\end{cases}
    \end{equation}
For $l\in \mathbb N$, we 
{correspondingly}
obtain the following supersymmetric partner Hamiltonians
\begin{eqnarray}\label{H1H2}
 \begin{cases}
    H_1=\omega_1A^\dagger_l A_l=-\frac{1}{2}\frac{d^2}{du^2}+\frac{l(l+1)}{2u^2}+\frac{1}{2}\omega_1^2u^2+\omega_1(l-\frac{1}{2})=H_{l}-\omega_1(l+\frac{3}{2}),\\\\
     H_2=\omega_1A_lA^\dagger_l=-\frac{1}{2}\frac{d^2}{du^2}+\frac{(l+1)(l+2)}{2u^2}+\frac{1}{2}\omega_1^2u^2-\omega_1(l+\frac{1}{2})=H_{l+1}-\omega_1(l+\frac{1}{2}).
 \end{cases}
\end{eqnarray}
These two Hamiltonians have the same energy
spectrum except the ground state of $H_2$
\begin{eqnarray}
 \begin{cases}
    H_1\Psi_{n}^{(l)}=\left[E_n^{(l)}+\omega_1-(l+\frac{3}{2})\right]\Psi_n^{(l)},\\\\
    H_2\Psi_{n+1}^{(l+1)}=\left[E_{n+1}^{(l+1)}-\omega_1(l+\frac{1}{2})\right]\Psi_{n+1}^{(l+1)}=\left[E_n^{(l)}-\omega_1(l+\frac{1}{2})\right]\Psi_{n+1}^{(l+1)}.
 \end{cases}
\end{eqnarray}
It is seen that these equations refer to the shape-invariance condition, by which we, equivalently, can write
\begin{eqnarray}
A^\dagger_{l-1}A_{l-1}- A_{l}A^\dagger_{l}=\frac{2}{\omega_1}.
\end{eqnarray}
Thus, altering the sequence of operators $A_l$ and $A^\dagger_l$ causes the value of $l$ to change. {This} 
 demonstrates how shape-invariance properties link the various factor orderings of the {Schr\"odinger--Wheeler--DeWitt} equation (\ref{sh2}). Shape-invariant potentials are well recognized for being simple to deal with when using lowering and raising operators, similar to the harmonic oscillator. However, we should note that the commutator of $A_l$ and $A^\dagger_l$ does not provide a constant value. Namely,
\begin{equation}
    [A_l,A^\dagger_l]=1+\frac{l+1}{\omega_1u^2}\,\,,
\end{equation}
which implies that these operators are not suitable to proceed with.
As {the eigenvalue relation (\ref{H1H2}) shows}, the potentials $V_1(u;b_0)$ and $V_2(u;b_1)$ introduced in (\ref{A-vs-W}) are given by
\begin{equation}
 \begin{split}
     V_1(u;b_0)&=\frac{b_0(b_0+1)}{2u^2}+\frac{1}{2}\omega_1^2u^2,\\
     V_2(u;b_1)&=\frac{b_1(b_1+1)}{2u^2}+\frac{1}{2}\omega_1^2u^2,
 \end{split}
\end{equation}
where $b_1=l+1$ and $b_0=l$. Therefore, in relation (\ref{trans-SI}) {this corresponds} to $\eta=1$.
According to the sub-section \ref{SIP}, we presume that replacing $l+1$ by $l$ in a given operator can be accomplished via a similarity transformation, (\ref{trans-op}), 
{and so we} build an appropriate algebraic structure via translation operator
\begin{eqnarray}\label{trans-op1}
T(l)=\exp\left( \frac{\partial}{\partial l}\right), \hspace{10mm}
T^{-1}(l)=T^{\dag}(l)=\exp\left(-\frac{\partial}{\partial l}\right).
\end{eqnarray}
Therefore, we introduce the  {operators}
\begin{eqnarray}
 B_l:={\frac{1}{\sqrt{2}}}T^\dagger(l)A_l,~~~B_l^\dagger:={\frac{1}{\sqrt{2}}}A^\dagger_lT(l),~~~ N^l_{\rm B}:=B_l^\dagger B_l,
\end{eqnarray}
which lead us to the simple harmonic oscillator (Heisenberg--Weyl) algebra
\begin{eqnarray}
 [B_l,B_l^\dagger]=1,~~~[ N^l_{\rm B},B_l]=-B_l,~~~[ N^l_{\rm B}, B_l^\dagger]=B_l^\dagger.
\end{eqnarray}
These commutation relations show that $B^\dagger_l$ and $B_l$ are the appropriate creation and annihilation
operators for the spectra of our shape-invariant potentials.
The action of these operators on normalized
eigenfunctions yields
\begin{eqnarray}
 \label{sh3}
 B_l\Psi_n^l=\sqrt{n}\Psi_{n-1}^l,~~~B_l^\dagger\Psi_n^l=\sqrt{n+1}\Psi_{n+1}^l,~~~ N^l_{\rm B}\Psi_n^l=n\Psi_n^l,~~~n=0,1,2,...~.
\end{eqnarray}
Equation (\ref{sh2}) and the last equation of the above set give:
\begin{eqnarray}
 E_n^l=\omega_1\left(2n+l+\frac{3}{2}\right).
\end{eqnarray}
In addition, the condition $ B_l\Psi_0^l=0$ gives us the ground state of the model universe for factor ordering $l$  by
\begin{equation}
    \Psi_0^l=C_l\exp{\left(-\frac{\omega_1}{2}u^2-\frac{l+1}{u^2}\right)},
\end{equation}
where $C_l$ is a normalization constant. The excited states can be {easily determined} by applying (\ref{gen-op-3}).

In what follows, let us complete our procedure by including the Grassmannian variables $\psi$ and $\bar{\psi}$, 
which satisfy
\begin{eqnarray}\label{psi}
\psi^2=0, \hspace{10mm} \bar{\psi}^2=0, \hspace{10mm}\psi\bar{\psi}+\bar{\psi}\psi=1,
\end{eqnarray}
{involving them into} the Hamiltonian \eqref{sham}. By means of such a procedure, we can {subsequently} construct a supersymmetric extension of our Hamiltonian: 
\begin{eqnarray}\label{s-Ham}
H_{\rm SUSY}=-\frac{1}{2}\frac{d^2}{du^2}+\frac{l(l+1)}{2u^2}+\frac{1}{2}\omega_1^2u^2+\omega_1\bar{\psi}\psi.
\end{eqnarray}
In our herein work, the convention of the left derivative for these variables has been adapted.
Up to now, we specified the bosonic creation and
annihilation operators $B$ and $B^\dagger$ in terms of the dynamical variable $u$ and its conjugate momenta. Here,
we can also introduce fermionic creation and
annihilation operators $C^{\dag}=\bar{\psi}$ and $C=\psi$. Therefore, the Hamiltonian \eqref{s-Ham} can {simply}
be written as
\begin{eqnarray}\label{s-Ham-2}
 H_{\rm SUSY}=2\omega_1(B^{\dag}B+C^{\dag}C).
\end{eqnarray}
 Adapting the basic commutator and anticommutator brackets
\begin{eqnarray}\label{bracket}
 [B,B^{\dag}]=1, \hspace{10mm} \{C,C^{\dag}\}=1,
\end{eqnarray}
and considering all the others to be zero, it is easy to show that the operators
\begin{eqnarray}\label{Oper}
 N_{\rm B}=B^{\dag}B, \hspace{10mm} N_{\rm F}=C^{\dag}C,\hspace{10mm} Q=B^{\dag}C,\hspace{10mm} Q^\dag=C^{\dag}B,
\end{eqnarray}
(where the indices $B$ and $F$ refer to the bosonic
and fermionic quantities, respectively) will be conserved quantities. Namely,
\begin{eqnarray}\label{cons}
[Q,H_{\rm SUSY}]=[Q^\dag,H_{\rm SUSY}]=0,\hspace{10mm}[ N_{\rm B},H_{\rm SUSY}]=[N_{\rm F},H_{\rm SUSY}]=0.
\end{eqnarray}
 Moreover, we have
\begin{eqnarray}\nonumber
 [Q,N_{\rm B}]&=&-Q, \hspace{10mm}  [Q,N_{\rm F}]=Q\\\nonumber
[Q^\dag,N_{\rm B}]&=&Q^\dag,    \hspace{10mm}   [Q^\dag,N_{\rm F}]=-Q^\dag,\\
\omega_1\{Q,Q^\dag\}&=&H_{\rm SUSY},\hspace{5mm} Q^2=\frac{1}{2}\{Q,Q\}=0,\hspace{3mm} \left(Q^\dag\right)^2=\frac{1}{2}\{Q^\dag,Q^\dag\}=0.
\label{algebra}
\end{eqnarray}
{From} \eqref{cons} and \eqref{algebra} an explicit algebra
 is {therefore produced}, where the Hamiltonian $H_{\rm SUSY}$ is a Casimir operator for the whole algebra \cite{Kumar:2011uz}.
If we use {a} matrix representation, then we can write,
{alternatively}
\begin{eqnarray}
 \psi=C=\begin{pmatrix}
 0&0\\
 1&0
 \end{pmatrix}, ~~~\bar\psi= C^\dagger=\begin{pmatrix}
 0&1\\
 0&0
 \end{pmatrix},
\end{eqnarray}
\begin{eqnarray}
 H_{\rm SUSY}=2\omega_1\begin{pmatrix}
 B^\dagger_{l} B_{l}+1&0\\
 0&B_lB_l^\dagger
 \end{pmatrix}=\begin{pmatrix}
 H_1&0\\
 0&H_2
 \end{pmatrix}.
 \end{eqnarray}

\section{Discussion}
\label{disc}
\indent

This paper is a review that embraces a twofold endorsement. On the one hand, it imports  SUSY features (in a quantum mechanical setting). However,  the current fact is that SUSY has not (yet…) been found in nature;  searches prevail for any evidence, being it  directly or indirectly. On the other hand, this review refers to quantum cosmology as a phenomenological domain regarding the full quantization of gravity.  Likewise, there is yet no clear-cut observational evidence of such a stage in the very early universe. Whereas the latter is fairly expected {as} the cosmos is further probed,  proceeding gradually  to prior times, the former, {although} alluringly  elegant, may just be a formal framework. So, producing a review on a topic involving these two ideas may seem twice likely to raise discomfort. But maybe not; perhaps SUSY quantum cosmology deserves to be kept nearby,  just  over an arm’s length, 
so to say\footnote{"Ah, but a man's reach should exceed his grasp, Or what's a heaven for?", Robert Browning (in 'Andrea del Sarto' l. 97 (1855))}, if the occasion (or data) emerges to either support it or at least enthuse more research about it. {There} are still  open aspects to appraise and the one brought on this review is among them
\cite{Moniz-1,Moniz-2}.

In what concerns SUSY quantum cosmology, there have been a few books in the  past 30 years or so  \cite{d2005supersymmetric,Moniz-1,Moniz-2} plus selected reviews on the different procedures that were constructed and
subsequently promoted
\cite{moniz1996supersymmetric,moniz2014quantum,garcia2021topics,moniz2014supersymmetric,lopez2015supersymmetric,obregon1998dirac,bene1994supersymmetric,csordas1995supersymmetric,kleinschmidt2009supersymmetric,macias1998supersymmetric,damour2013quantum}. 
Likewise  chapters (and sections) about SUSY quantum cosmology in well known textbooks {concerning quantum gravity}
\cite{Kiefer:2004xyv,esposito2009quantum,calcagni2017classical,bojowald2011quantum}. In particular, the direction and extension of $N=2$ SQM to SUSY quantum cosmology 
was led by \cite{lidsey1995quantum,lidsey2000supersymmetric,graham1991supersymmetric,lidsey1995scale} and subsequently by  \cite{tkach1996supersymmetric,obregon1996superfield},  referring to conformal issues. 

Therefore, the opportunity to produce this review enthused us to refer and to explore a specific particular aspect that was  indicated as a concrete open problem  {in} SUSY quantum cosmology (Nb. {we emphasize} many more  still remain; cf. in \cite{Moniz-1,Moniz-2}). {Concretely, investigating the setting of SIP, fairly present in $N=2$  SQM. This  framework  has been developed  {previously} and independently from SUSY quantum cosmology, to explore issues in supersymmetric quantum field theory, namely SUSY breaking (which the seminal paper \cite{witten1981dynamical} has made possible). }

Hence, a necessary and yet to be done analysis
%in this context 
{remained} to be elaborated: explicitly considering SIP as we just mentioned {within quantum cosmology}. I.e., bringing up this possibility and using it intertwined within {SUSY 
quantum cosmology}. {Thus} is a new idea 
for other researchers to pick up 
%on our review 
{and} evolve  forward,  producing their own assertions. The content of our review is thus very open, 
%and it is the setting of 
{conveying} a direction {for further exploration}.
{It involves} algebraic quantum-mechanical {aspects} that are present when SIP characterizes particular models. {It also 
deals with} integrability, which  SUSY seems to bring so elegantly.
%, and where SIP is an additional attractive feature to consider furthermore in SUSY quantum cosmology.  

{In this paper, besides  contributing with a topical review towards this Special Issue, we also provided 
a constructive example to illustrate how promising the framework can be. 
Concretely, we 
provided a case study, consisting of  a spatially flat FRW model in the presence of a single scalar field, 
minimally coupled to gravity. We 
extracted the 
Schr\"odinger--Wheeler--DeWitt equation 
containing a  particular set
of possible factor ordering. Next, we
computed the corresponding supersymmetric partner Hamiltonians. Intriguingly,   the shape invariance 
properties can be related to  the several factor orderings of our Schr\"odinger--Wheeler--DeWitt equation. The 
ground state was 
computed and the excited states as well. Consistently, 
 the partner  
Hamiltonians,
were  explicitly presented within a $N=2$ 
SQM framework.} 

{We made implicitly another suggestion  in  Section  \ref{int} (Introduction). In more detail, we suggested  to build a twofold framework.   On the one hand, importing  the ideas employed in references} \cite{Shahram-1,Shahram-2,Shahram-3, Shahram-4,Shahram-5}, where the presence of constraints, their algebra plus a natural integrability induces  separability in the Hilbert space of solutions for the Wheeler--DeWitt equation. {On the other hand,} 
explore, at least to begin with  on formal terms, whether any such algebra of the constraints  generators for a minisuperspace, would bear any   similarity to an algebra of supersymmetry generators. 
In others words, perhaps producing a sequence of 
\textit{new} operators $A_l$ and $A^\dagger_l$, assisting
SIP  properties but also related to SUSY constraints  of a Schr\"odinger--Wheeler--DeWitt equation similar to (\ref{sh2}).
{SIP} are well recognized for being simple to deal. 
In essence, {our suggestion is to explore ($i$) if} there is any relation between SIP and the descriptive report in \cite{Shahram-1,Shahram-2,Shahram-3, Shahram-4,Shahram-5} and,
if positive, ($ii$)  apply it within SUSY quantum cosmology.

\section*{Acknowledgments}

PVM and SMMR acknowledge the FCT grants UID-B-MAT/00212/2020 and UID-P-MAT/00212/2020 at CMA-UBI plus
%COST Action CA15117 (CANTATA) 
the COST Action CA18108 (Quantum gravity phenomenology in the multi-me\-ssen\-ger approach).

\bibliographystyle{unsrt}
\bibliography{Shape}

\end{document}